\documentclass[twocolumn,amsmath,amssymb,superscriptaddress,prb,a4paper]{revtex4-1}
\usepackage{graphicx}
\usepackage[paperwidth=212mm,paperheight=297mm,centering,hmargin=1.65cm,vmargin=2.6cm]{geometry}

\usepackage{graphicx}                       
\usepackage{hyperref}                       
\usepackage{dsfont}
\usepackage{mathrsfs}
\usepackage{bigints}
\usepackage{bbold}
\usepackage{amsthm}
\usepackage{setspace}

\usepackage{xcolor,colortbl}

\newcommand{\ketbra}[2]{\left|#1\right\rangle\hskip-1mm\left\langle #2\right|}

\newcommand{\ket}[1]{\vert#1\rangle}
\newcommand{\bra}[1]{\langle#1\vert}
\newcommand{\outter}[1]{\ket{#1}\bra{#1}}

\newcommand{\mcL}{\mathcal{L}}

\newcommand{\beq}{\begin{eqnarray}}
\newcommand{\eeq}{\end{eqnarray}}

\newcommand{\tr}{\operatorname{tr}}

\usepackage{array}
\newcolumntype{L}[1]{>{\raggedright\let\newline\\\arraybackslash\hspace{0pt}}m{#1}}
\newcolumntype{C}[1]{>{\centering\let\newline\\\arraybackslash\hspace{0pt}}m{#1}}
\newcolumntype{R}[1]{>{\raggedleft\let\newline\\\arraybackslash\hspace{0pt}}m{#1}}

\begin{document}
\title{Phonon limit to simultaneous near-unity efficiency and indistinguishability in semiconductor single photon sources}

\author{Jake Iles-Smith}
\thanks{These authors contributed equally to this work}
\affiliation{Department of Photonics Engineering, DTU Fotonik, {\O}rsteds Plads, 2800 Kongens Lyngby, Denmark}
\affiliation{{Photon Science Institute \& School of Physics and Astronomy, The University of Manchester, Oxford Road, Manchester M13 9PL, United Kingdom}}
\author{Dara P. S. McCutcheon}
\thanks{These authors contributed equally to this work}
\affiliation{Quantum Engineering Technology Labs, H. H. Wills Physics Laboratory and Department of Electrical and Electronic Engineering, 
University of Bristol, BS8 1FD, UK}
\author{Ahsan Nazir}
\affiliation{{Photon Science Institute \& School of Physics and Astronomy, The University of Manchester, Oxford Road, Manchester M13 9PL, United Kingdom}}
\author{Jesper M\o{}rk}
\affiliation{Department of Photonics Engineering, DTU Fotonik, {\O}rsteds Plads, 2800 Kongens Lyngby, Denmark}
\date{\today}
\pacs{Valid PACS appear here}

\begin{abstract}
Semiconductor quantum dots have recently emerged as a leading platform to efficiently generate highly indistinguishable 
photons, and this work addresses the timely question of how good these solid-state sources can ultimately be. 
We establish the crucial role of lattice relaxation in these systems in giving rise to 
trade-offs between indistinguishability and efficiency. 
We analyse the two source architectures most commonly employed: 
a quantum dot embedded in a waveguide and a quantum dot coupled to an optical cavity. 
For waveguides, we demonstrate that the broadband Purcell effect results in a simple inverse relationship, where indistinguishability 
and efficiency cannot be simultaneously increased. 
For cavities, the frequency selectivity of the Purcell enhancement results in a more subtle trade-off, 
where indistinguishability and efficiency can be simultaneously increased, though by the same mechanism not arbitrarily, 
limiting a source with near-unity indistinguishability ($>99$\%) to an efficiency of approximately 96\% for realistic parameters.
\end{abstract}
\maketitle

The efficient generation of on-demand highly indistinguishable photons remains a barrier to the scalability of 
a number of photonic quantum technologies~\cite{OBrien2009,Knill2001,RevModPhys.79.135,Pan2012}. 
To this end, attention has recently turned towards solid-state systems, 
and in particular semiconductor quantum 
dots (QDs)~\cite{Michler22122000,Santori2001,Santori2002,gazzano13,Nowak2014,He2013,Thoma2016,Somaschi2016,Ding2016}, 
which can not only emit a single photon with high quantum efficiency, but 
can be easily integrated into larger photonic structures~\cite{lodahl2015interfacing}, 
resulting in photons being emitted into a well-defined mode and direction. 
Highly directional emission is crucial to the overall efficiency of the source, and is typically achieved by 
either placing the QD in a waveguide with low out-of-plane scattering~\cite{Claudon2010,PhysRevLett.113.093603}, or by coupling resonantly to 
an optical cavity mode~\cite{Somaschi2016,Ding2016,Santori2001,Santori2002,gazzano13,Nowak2014}. 
Nevertheless, the solid-state nature of QDs leads to strong coupling 
between the electronic degrees of freedom and their local environment; 
fluctuating charges~\cite{Houel2012}, nuclear spins~\cite{Kuhlmann2013,Kuhlmann2015}, and 
lattice vibrations~\cite{PhysRevLett.104.017402,PhysRevLett.105.177402,McCutcheon2013,nazir2015modelling} all lead to a suppression 
of photon coherence and a resulting reduction in 
indistinguishability~\cite{PhysRevB.87.081308,PhysRevB.90.035312,Thoma2016,iles2016fundamental,Unsleber2015}. 
While early experiments were indeed limited by these 
factors~\cite{Santori2001,Santori2002,gazzano13,Nowak2014}, improvements in fabrication and resonant excitation 
techniques have steadily increased photon indistinguishability to levels now exceeding 99\% in resonantly coupled 
QD--cavity systems~\cite{Somaschi2016,Ding2016}. 
Photon extraction efficiencies have also steady improved, with the highest values reaching $98\%$ in 
a photonic crystal waveguide~\cite{PhysRevLett.113.093603}. 

Despite this impressive progress, a system boasting very high ($>99\%$) indistinguishability {{and}} 
efficiency as required for e.g. cluster state quantum computing~\cite{Aharonovich2016} remains elusive. 
Strategies aimed at achieving such a source typically focus on engineering the photonic environment in order 
to maximise the Purcell effect~\cite{Purcell1946,Rao2007}, 
where the QD emission rate becomes $F_P\Gamma$, 
with $\Gamma$ the bulk emission rate and $F_P$ the Purcell factor~\cite{Purcell1946}. 
Modelling a QD as a simple two-level-system with a Markovian phenomenological dephasing rate $\gamma$, 
the Purcell factor allows one to quantify the indistinguishability and efficiency as 
$\mathcal{I}=\Gamma F_P/(\Gamma F_P+2\gamma)$ and $\eta = F_P/(F_P+1)$ respectively~\cite{Kaer2013,Bylander2003}.
In this simplistic model, one concludes that the Purcell factor is the key quantity of interest, 
which when increased will simultaneously lead to greater indistinguishability and efficiency. 

In this work we demonstrate that this reasoning fails
when one considers the coupling of the QD to its solid-state lattice at a microscopic level. 
We show that even in an idealised scenario, in which all other sources of noise are suppressed, 
the unavoidable coupling to phonons means neither waveguide nor cavity based sources can simultaneously 
reach near-unity indistinguishability and efficiency through Purcell enhancement alone. \\

\section*{Results}

In contrast to simply introducing a Markovian dephasing rate, exciton--phonon coupling in the QD causes the 
lattice to adopt different configurations depending on whether the QD is in its ground or excited state [see Fig.~1]. 
As such, an excited to ground state transition accompanied by photon emission into the zero phonon line (ZPL)
has a probability which scales as the square of the Franck--Condon factor $B< 1$, 
corresponding to the overlap of the two lattice configurations. The remaining emission 
events also scatter phonons in the process, resulting in emission of distinguishable photons, 
and a phonon sideband (SB) in the spectrum which must be removed. Due to the broadband nature 
of the Purcell enhancement in waveguides, the SB can only be removed by filtering. 
This necessarily sacrifices efficiency, resulting in a simple trade-off between indistinguishability and efficiency. 
For an emitter embedded in a moderate to high Q-cavity the phonon sideband can be naturally suppressed, 
though in this case the efficiency becomes 
 $\eta=B^2 F_P /(B^2 F_P+1)$, showing that removal of the sideband reduces the expected efficiency through 
 the Franck--Condon factor. This can in part be compensated by increasing the Purcell enhancement, 
 though not indefinitely, as both the efficiency and indistinguishability drop when the strong coupling regime is reached. 
Based on a rigorous non-Markovian phonon 
theory, we derive analytic results quantifying the performance of single-photon sources for different 
architectures and in different regimes of operation. 

\begin{figure}
\begin{center}
\includegraphics[width=0.48\textwidth]{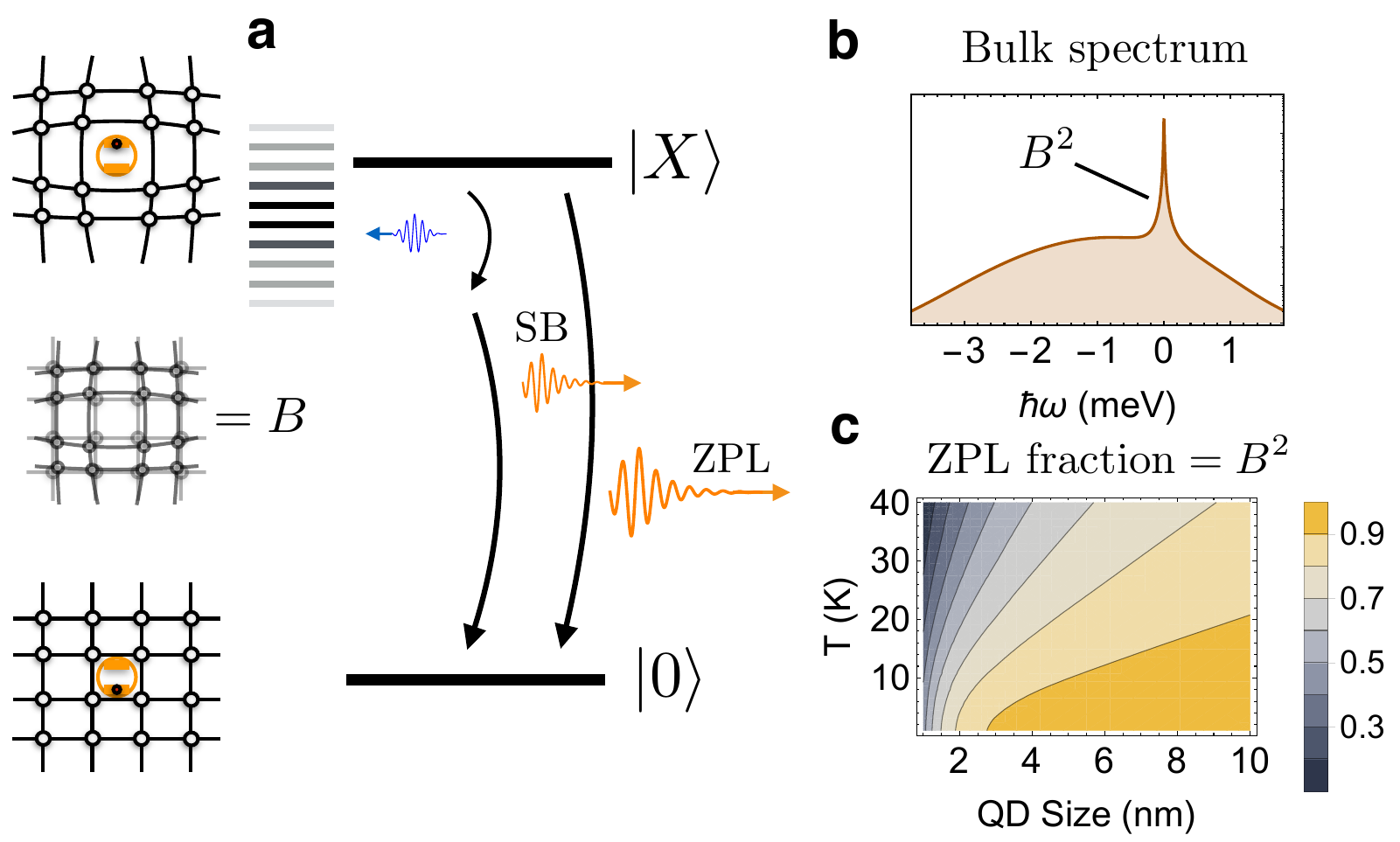}
\caption{Lattice relaxation in quantum dots and the phonon sideband. a, The ground state $\ket{0}$ 
and single exciton state $\ket{X}$ of a quantum dot are each associated with 
a different lattice configuration, shown on the left. b, Probability of emission into the zero phonon line (ZPL) scales as the square of the Franck--Condon 
factor $B$ (lattice wavefunction overlap), with the remaining emission events constituting the phonon sideband (SB). c, The fraction of emission 
into the ZPL decreases with temperature and increases with quantum dot size.}
\end{center}
\end{figure}

\subsection*{Phonon interactions in optically active QDs}
The two key quantities used to characterise a single photon source are the 
efficiency, defined as~\cite{Kaer2013,Kiraz2004}
\begin{equation}
\eta = \frac{P_D}{P_D+ P_O },
\label{EfficiencyDefinition}
\end{equation}
and the photon indistinguishability, defined as
\begin{equation}
\mathcal{I} = P_D^{-2}\int_{-\infty}^{\infty}d\omega\int_{-\infty}^{\infty}d\nu~|S_D(\omega,\nu)|^2,
\label{IndistinguishabilityDefinition}
\end{equation}
where the $D$ and $O$ subscripts denote the detected field and the field lost into unwanted modes. 
Here $S_{D,O}(\omega,\nu) = \langle E_{D,O}^{\dagger}(\omega)E_{D,O}(\nu)\rangle$ is the generalised 
two-colour spectrum, with $E_{D,O}(\omega)$ the positive 
component of the electric field in frequency space. 
Eq.~({\ref{IndistinguishabilityDefinition}}) is more commonly (and equivalently) written~\cite{Kaer2013} 
$\mathcal{I}=P_D^{-2}\int_0^{\infty} d t\int_0^{\infty}d\tau |\langle \tilde{E}^{\dagger}(t+\tau)\tilde{E}(t)\rangle|^2$, 
where $\tilde{E}_{D,O}(t)=\smash{\int_0^{\infty}d t \mathrm{e}^{-i \omega t} E_{D,O}(\omega)/(2\pi)}$. 
For $\omega=\nu$ the two-colour spectrum is the measured emission spectrum, 
and the power into each channel is  
$P_{D,O} = \smash{\int_{-\infty}^\infty d\omega S_{D,O}(\omega,\omega)}$. 
These expressions highlight the essential connection between the spectrum and performance of the source. 
We will analyse the three commonly used single photon source architectures shown in Fig.~2~(a); 
a QD in a waveguide with Purcell enhancement (a slow-light waveguide) 
without (i) and with (ii) a spectral filter, and a QD coupled to a cavity (iii). 

Calculation of the source figures of merit requires an accurate model of the dephasing processes affecting the QD. 
In addition to phonon induced processes, charge noise and spin noise can also affect emitted 
photon coherence~\cite{Kuhlmann2013,Kuhlmann2015}. 
However, our purpose here is assess the ultimate limits of a QD based source, 
and note that charge and spin noise can be heavily suppressed 
in suitably engineered samples~\cite{Somaschi2016,Ding2016}, 
while coupling to phonons can ever be completely quenched, 
as even at $T=0~\mathrm{K}$ phonon emission can still take place. 
We therefore focus on phonon induced dephasing mechanisms, 
with the understanding that our numerical results  
correspond to best case scenarios. 
Nevertheless, due to the very fast timescale ($\sim\mathrm{ps}$) 
associated with phonon relaxation compared to the other dephasing mechanisms mentioned above, 
charge and spin noise can be readily included 
within our formalism by the introduction of Markovian dephasing rates, and our analytical 
expressions will explicitly include these rates also.

Of the possible phonon interactions that can take place in QDs, coupling to longitudinal acoustic (LA) 
phonons via deformation potential coupling has been shown to dominate~\cite{PhysRevLett.105.177402,PhysRevLett.104.017402}.
Aside from lattice relaxation as captured by the Franck--Condon factor mentioned above, 
above a certain temperature 
LA phonons can also induce virtual transitions to QD states beyond the lowest single exciton state, giving rise 
to an additional phonon mediated decoherence process quite different in nature to the real phonon transitions 
represented by the emission spectrum sideband~\cite{Muljarov2004}. These processes are expected 
to be heavily suppressed at low temperatures $(T<10~\mathrm{K})$, and will therefore be neglected 
in what follows, though once again we note that their inclusion could be easily achieved owing to the 
drastically different timescales involved. 

\begin{figure*}
\begin{center}
\includegraphics[width=0.99\textwidth]{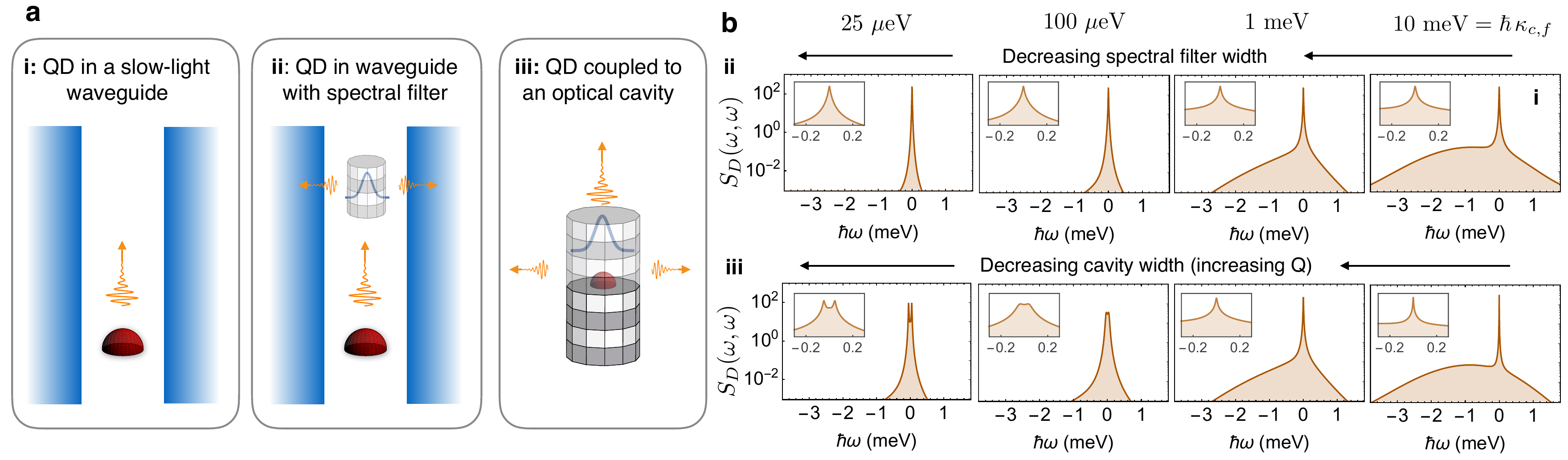}
\caption{Single photon source architectures and emission spectra. a, (i)--(iii) show the architectures we analyse: 
a QD emitting into a slow-light waveguide 
with and without a spectral filter, and a QD in a coherently coupled optical cavity. 
b, (ii)--(iii) show corresponding emission spectra as the filter or cavity is reduced 
in spectral width, demonstrating the filtering property of a cavity. The insets show a zoom-in 
of the ZPL features, highlighting ZPL broadening (Purcell enhancement) in the cavity case, which 
ultimately gives rise to vacuum Rabi splitting.
The unfiltered spectrum for case (i) closely resembles the 
broad filter $\hbar\kappa_f=10~\mathrm{meV}$ case in (ii) as indicated. 
Parameters: $T=4~\mathrm{K}$, $\alpha=0.03~\mathrm{ps}^2$, 
$\hbar\xi=1.45~\mathrm{meV}$, $\hbar\Gamma=1~\mu\mathrm{eV}$; the 
waveguide in (i) and (ii) has Purcell factor $\Gamma_D/\Gamma=10$, 
while the cavity in (iii) has $\hbar g=50~\mu\mathrm{eV}$, giving 
$\Gamma_{\mathrm{cav}}/\Gamma=10$ when $\hbar\kappa_c=1~\mathrm{meV}$.}
\end{center}
\end{figure*}

With these arguments in mind, we consider a QD as a two-level-system with ground state $\ket{0}$ 
and single exciton state $\ket{X}$ with energy 
$\hbar\,\omega_X$~\cite{PhysRevLett.105.177402,PhysRevLett.114.137401,PhysRevB.90.035312,bimberg1999quantum,zrenner02,1367-2630-12-11-113042}. 
The QD is coupled to a phonon and photon environment, giving the Hamiltonian 
$H = \hbar\,\omega_X\ket{X}\bra{X} + \smash{H_{I}^{PH}+H_{I}^{EM}  + H_E^{PH}+H_E^{EM}}$, 
where $H_E^{PH}$ and $H_E^{EM}$ describe the free evolution of the phonon and photonic environments. 
The term $H_I^{EM}$ contains the electric field operators $E_{D,O}(\omega)$ which determine the spectrum, 
and describes the interaction between the QD and its photonic environment. 
Coupling to LA phonons is captured by the terms~\cite{PhysRevLett.105.177402,PhysRevLett.114.137401}, 
$\smash{H_{I}^{PH} = \hbar\ket{X}\bra{X}\sum_k g_k(b_k^\dagger + b_k)}$ and 
$H_E^{PH}=\hbar\sum_k \nu_k b_k^{\dagger}b_k,$ where 
$b_k$ ($\smash{b_k^\dagger}$) is the annihilation (creation) 
operator of the phonon mode with wavevector $k$ and frequency $\nu_k$. 
This interaction captures the mechanical deformation of the lattice 
when an exciton is present in the QD [see Fig.~(1)]. 
Despite the complexity of the QD--phonon interaction, the harmonic nature of 
the phonons means their interaction with the QD can be fully characterised by the phonon spectral density, 
which for a spherically symmetric QD with harmonic confinement potential can be written~\cite{nazir2008photon,nazir2015modelling}
$J_{\mathrm{ph}}(\nu) = \sum_k \vert g_k\vert^2\delta(\nu-\nu_k) = \alpha\nu^3 \exp[-\nu^2/\xi^2]$. 
Here $\alpha$ is an overall exciton--phonon coupling strength, and $\xi=\sqrt{2} v/d$ 
is the phonon cut-off frequency, with 
$v$ the speed of sound and $d$ the confinement length (QD size). The cut-off frequency $\xi$ 
defines a phonon energy scale above which interactions with the exciton are 
suppressed due to a mismatch in phonon and QD length scales.

Though the Hamiltonian given above, together with an appropriate choice of $H_I^{EM}$ 
to model the relevant photonic environment, completely specifies the problem, calculating the 
two-colour spectra $S_{D,O}(\omega,\nu)$ and by extension the source figures of 
merit is extremely challenging. 
In general the Hamiltonian is not easily diagonalised, and typically one therefore 
turns to approximate methods from the theory of open quantum systems, 
for example perturbative Markovian approaches such as the time-convolutionless master equation 
technique~\cite{PhysRevB.90.035312,PhysRevB.87.081308}. 
Since the emission spectrum sideband results from changes to the phonon environment (lattice relaxation), 
it is non-Markovian in nature~\cite{mccutcheon2015optical}, and as such these Markovian treatments fail to capture it, 
yielding inaccurate source figures of merit~\cite{PhysRevB.90.035312,PhysRevB.87.081308}. 
Non-Markovian master equations can be employed~\cite{PhysRevB.90.035312}, 
though using these to calculate spectra requires extensions to the quantum regression 
theorem~\cite{mccutcheon2015optical}, which had limited success when used to 
calculate photon indistinguishability, giving results that appeared not to approach 
the known analytic result in the limit of no cavity or filtering effects~\cite{PhysRevB.90.035312}. 
To date brute force numerical approaches, based on exact diagonalisation~\cite{Kaer2013,PhysRevB.90.035312} or 
non-equilibrium Green's functions techniques~\cite{Grange2016} have had 
the most success, though these provide 
limited insight into the underlying physical processes 
involved, and only in rare cases give analytic expressions. 

To overcome these difficulties, we adopt a polaron transform approach, used in conjunction with formally 
solving the Heisenberg equations of motion for the emitted fields. This allows the dominant 
non-perturbative non-Markovian phonon influence to be included, 
and permits us to derive analytic expressions in relevant regimes which elucidate the interplay between the 
Purcell and Franck--Condon factors, and trade-offs between efficiency and indistinguishability. 
Full details of the polaron transformation are given in the Supplementary information, 
though the central idea is to apply a displacement to the phonon mode operators dependent 
on the QD state, $b_k\to b_k-\ketbra{X}{X}g_k/\nu_k$, as this removes 
the original exciton--phonon coupling from the 
Hamiltonian~\cite{1367-2630-12-11-113042,Roy2011,nazir2015modelling,PhysRevB.92.205406,Iles-Smith:16}. 
Unitarity of the mode displacement means 
that the QD states must transform as 
$\ket{0}\to\ket{0}$ and $\ket{X}\to B_+\ket{X}$ with $\smash{B_+ = \exp[\sum_k \nu_k^{-1}g_k(b_k^\dagger - b_k)]}$, 
and we can identify $B_+$ as the operator achieving 
the necessary displacement of the lattice associated with the presence of an exciton. 
The Franck--Condon factor is then the thermal expectation value of this 
lattice displacement operator: 
\begin{equation}
B =\langle B_{+} \rangle= \exp\bigg[-\frac{1}{2}\int_{0}^\infty d\nu \frac{J(\nu)}{\nu^2}\coth\Big(\frac{\hbar\,\nu}{2 k_B T}\Big)\bigg]. 
\end{equation} 
As mentioned, with no cavity or filtering effects only $B^2$ of photon emission events 
go into the ZPL, with the remainder being incoherent in nature 
and constituting a phonon SB in emission spectra. 
As seen in Fig.~1~(b), while this phonon SB is orders of magnitude lower in intensity, 
its width is determined by the phonon cut-off frequency $\hbar\xi\sim 1~\mathrm{meV}$ for typical parameters. 
As such, even at $T=0$~K where only phonon emission occurs, the sideband constitutes $\approx7\%$ of the emission, 
which increases with temperature and for QDs with smaller exciton localisation lengths, 
as seen in Fig.~1~(c). 

\begin{figure}
\center
\includegraphics[width = 0.48\textwidth]{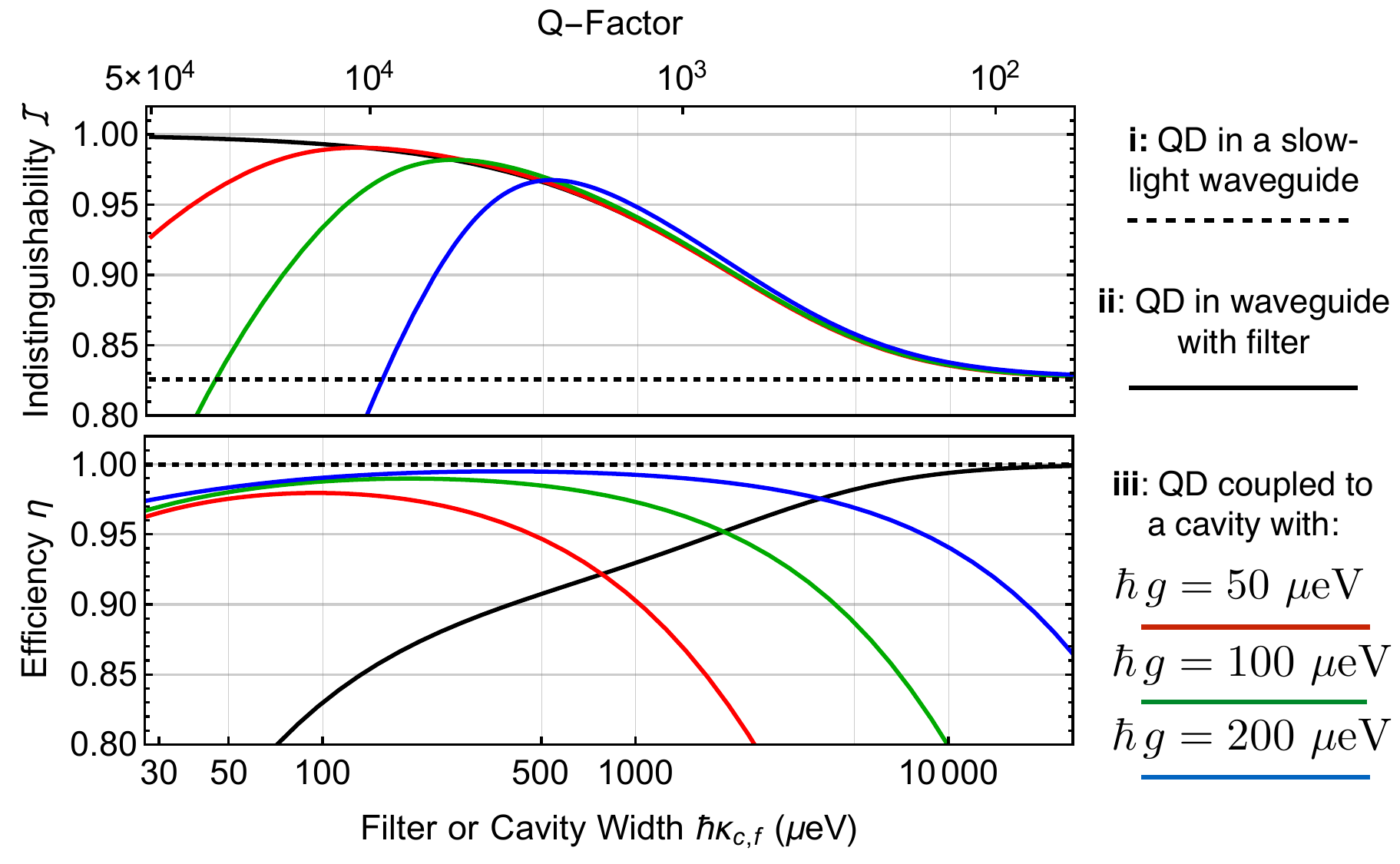}
\caption{Indistinguishability and efficiency 
of the three source setups shown in Fig.~(2).  
The indistinguishability plot demonstrates that the dominant effect of a resonantly coupled cavity is to filter the QD emission, 
while the efficiency plot demonstrates that Purcell enhancement in a cavity can overcome efficiency losses incurred by 
filtration of the phonon sideband. Cavity Q-factors on the upper x-axis correspond to a cavity resonance $\hbar\omega_c=1.4~\mathrm{eV}$. 
Parameters as in Fig.~(2), giving $B^4=83\%$, and we have assumed a loss-less 
waveguide $\Gamma_O=0$ for (i) and (ii), while for (iii) $\Gamma_O=\Gamma$.
}
\end{figure} 

\subsection*{Emission properties}
Our task now is to understand how a spectral filter or cavity can affect the 
detected spectrum $S_{D}(\omega,\nu)$, which will in turn affect the indistinguishability 
via Eq.~({\ref{IndistinguishabilityDefinition}}) by, for example, removing the phonon SB. 
Crucially, however, we also need 
to understand the quantitative relationship between the detected and lost (out-of-plane) 
spectrum $S_{O}(\omega,\nu)$ when these filtering or cavity affects are introduced, since 
this will affect the source efficiency via Eq.~({\ref{EfficiencyDefinition}}). 

As shown in Methods, the two-colour-spectra are found by 
solving the Heisenberg equations of motion for the electric field operators, and 
in all cases (i)--(iii) we find it is possible to write $S_{D}(\omega,\nu) = \mathcal{G}(\omega,\nu) S_{O}(\omega,\nu)$. 
The function $\mathcal{G}(\omega,\nu)$ is a Green's function, describing   
how the field is transformed propagating from its creation at the QD, to the detector. 
For the unfiltered waveguide source (i) $\mathcal{G}(\omega,\nu)=\smash{\Gamma_D/\Gamma_O}$, with 
$\Gamma_D$ and $\Gamma_O$ the emission rates into and out of the waveguide, 
showing that the in-plane spectrum is simply a frequency independent enhancement 
of the out-of-plane spectrum.
For the filtered waveguide source (ii) 
$\mathcal{G}(\omega,\nu)=\smash{(\Gamma_D/\Gamma_O)h_f^*(\omega)h_f(\nu)}$, where 
$h_f(\omega) = (\kappa_f/2)[i(\omega-\omega_f)-(\kappa_f/2)]^{-1}$ with 
$\kappa_f$ and $\omega_f$ the filter width and central frequency respectively. 
The filter now fundamentally changes the 
detected spectrum, as we might expect. 
As a key insight of this work, in case (iii) for the optical cavity we find 
$\mathcal{G}(\omega,\nu)=\smash{(\Gamma_{\mathrm{cav}}/\Gamma_O)h_c^*(\omega)h_c(\nu)}$, where now 
$h_c(\omega) = i(\kappa_c/2)[i(\omega - \omega_{c} )-\kappa_c/2]^{-1}$ and $\Gamma_{\mathrm{cav}}=4 g^2/\kappa_c$, 
with $g$ the light--matter coupling strength, 
$\kappa_c$ the cavity width, and $\omega_c$ the cavity mode frequency. 

Comparing cases (ii) and (iii) above, we see that there is a formal analogy 
between a spectral filter and an optical cavity, as has been alluded to elsewhere~\cite{PhysRevB.92.205406,Grange2015}. 
That is not to say, however, that the two are equivalent; 
as a filter is reduced in width and the sideband removed, one simply moves 
photons from the detected channel to the out-of-plane channel. 
As a cavity is reduced in width, however, the strength of the light--matter coupling is modified, 
giving rise to Purcell enhancement of emission events resonant with the cavity, while 
also removing the sideband.  
Unlike a filter, this cavity enhancement can overcome sideband photons that are now being 
lost due to cavity filtering effects. 
The broadband nature of Purcell enhancement in waveguides 
means a waveguide with Purcell enhancement and a filter is not equivalent to a 
cavity, since in the former case both the ZPL and the sideband are enhanced.  
Detected spectra for the waveguide with filter (ii) and cavity (iii) are shown in Fig.~2~(b), 
where the waveguide has a Purcell factor of $\Gamma_D/\Gamma=10$. In addition to 
filtering effects seen in both cases, the insets show that in the cavity case, 
frequency selectivity of the Purcell enhancement gives ZPL broadening, and ultimately signs of vacuum 
Rabi splitting as the strong coupling regime is reached. 

\subsection*{Waveguide vs Cavity Comparison}
In Fig.~(3) we compare the three single photon source architectures shown in Fig.~2~(a). 
For large cavity or filter widths ($\kappa_{c,f}\gg \xi\sim1~\mathrm{meV}/\hbar$), the entire sideband contributes to the 
detected field [see Fig.~2~(b)], yielding an indistinguishability of that in bulk, 
$\mathcal{I}=B^4\approx 83\%$ for realistic parameters at $T=4~\mathrm{K}$. 
As the filter or cavity is reduced in width, the indistinguishability increases as the phonon sideband is removed. This plot  
demonstrates that until the strong coupling regime is reached, i.e. for $\kappa_c>4g$, with regards to the indistinguishability, 
the dominant effect of the cavity is that of filtering, as also suggested by Fig.~2~(b). 
The efficiency of the filtered source (ii), however, always decreases with decreasing filter width as the sideband is removed, 
whereas the cavity efficiency (iii) increases, since the Purcell effect compensates for photons lost into the sideband.

To elucidate these points, let us consider the experimentally relevant regime where the filter or cavity 
width is larger than any features present in the ZPL. This corresponds to $\Gamma_D<\kappa_f$ in case (ii), and 
$\Gamma_{\mathrm{cav}}<\kappa_c$ in case (iii), meaning that the strong coupling regime is not reached. In this regime 
we find that the master equation describing the QD degrees of freedom can be approximated as
$\dot{\rho}=\Gamma_{\mathrm{tot}} \mathcal{L}_{\sigma}[\rho(t)]+2\gamma_{\mathrm{tot}} \mathcal{L}_{\sigma^{\dagger}\sigma}[\rho(t)]$, 
where for (i) and (ii) $\Gamma_{\mathrm{tot}}=\Gamma_O+\Gamma_D$ and $\gamma_{\mathrm{tot}}=\gamma$, 
and for case involving the cavity (iii) $\Gamma_{\mathrm{tot}}=\Gamma_O+\Gamma_{\mathrm{cav}}$ and 
$\gamma_{\mathrm{tot}}=\gamma+\gamma_{\mathrm{ph}}$ with 
$\gamma_{\mathrm{ph}}=2\pi(g B/\kappa_c)^2 J_{\mathrm{ph}}(2 g B)\coth(\hbar g B/k_B T)$ a 
Markovian phonon-induced ZPL dephasing rate. 
We have introduced a phenomenological dephasing rate $\gamma$ to capture e.g. charge or spin noise, 
which is a valid procedure provided it is uncorrelated with any phonon processes.
With this master equation we find that the indistinguishability can be approximated by 
\beq
\label{IWithSB}
\mathcal{I}=\frac{\Gamma_{\mathrm{tot}}}{\Gamma_{\mathrm{tot}}+2\gamma_{\mathrm{tot}}}\left(\frac{B^2}{B^2+\mathcal{F}[1-B^2]}\right)^2,
\eeq
where $\mathcal{F}=\int_{-\infty}^{\infty} d\omega |h_{f,c}(\omega)|^2 S_{\mathrm{SB}}(\omega,\omega)/\int_{-\infty}^{\infty} d\omega S_{\mathrm{SB}}(\omega,\omega)$
is the fraction of the sideband not removed by the filter or optical cavity. 
The first factor in Eq.~({\ref{IWithSB}}) is similar to the phenomenological expression~\cite{Kaer2013}, though with  
an additional phonon-induced dephasing rate $\gamma_{\mathrm{ph}}$. 
The second factor, however, 
highlights the essential role of the Franck--Condon factor $B$, and the 
interplay between this and the fraction of the sideband remaining in the spectrum $\mathcal{F}$. 
The efficiency in this regime is given by 
\beq
\eta = \frac{\Gamma_{\mathrm{cav}}(B^2+\mathcal{F}[1-B^2])}{\Gamma_{\mathrm{cav}}(B^2+\mathcal{F}[1-B^2])+\Gamma_O},
\eeq
for the cavity, and $\eta = (B^2+\mathcal{F}[1-B^2])\Gamma_D/(\Gamma_D+\Gamma_O)$ for the waveguide, 
again demonstrating the importance of the Franck--Condon factor. 

For a broad filter or low-Q cavity, for which $\kappa_{f,c}\gg \xi\sim 1~\mathrm{meV}/\hbar$, 
we have $\mathcal{F}=1$ and 
Eq.~({\ref{IWithSB}}) becomes $\mathcal{I}=B^4\Gamma_{\mathrm{tot}}/(\Gamma_{\mathrm{tot}}+2\gamma_{\mathrm{tot}})$. 
Since $B <1$, the phonon sideband reduces the indistinguishability that would be expected from Markovian 
or phenomenological treatments. The efficiencies in this regime become 
$\eta = \Gamma_D/(\Gamma_D+\Gamma_O)$ in the waveguide case, 
while for the cavity we find $\eta = \Gamma_{\mathrm{cav}}/(\Gamma_{\mathrm{cav}}+\Gamma_O)$, becoming  
$\eta=F_{\mathrm{cav}}/(F_{\mathrm{cav}}+1)$ for $\Gamma_O=\Gamma$ with $F_{\mathrm{cav}}=4 g^2/(\kappa_c\Gamma)$ the cavity Purcell factor. 
Thus, in this regime the efficiencies are equal to those expected from phenomenological approaches~\cite{Kaer2013}.

For a sufficiently narrow filter or cavity, for which
$\kappa_{f,c}\ll \xi$, we have $\mathcal{F}\approx 0$, and 
Eq.~({\ref{IWithSB}}) becomes $\mathcal{I}=\Gamma_{\mathrm{tot}}/(\Gamma_{\mathrm{tot}}+2\gamma_{\mathrm{tot}})$. 
Here the cavity or filter removes the phonon sideband from the detected spectrum, increasing the indistinguishability 
as compared to that found for a broad filter or low-Q cavity. 
Although the sideband appears not to affect the indistinguishability of the 
source in this regime, the efficiency drops monotonically in case (ii), and for the cavity (iii) becomes 
$\eta = B^2 \Gamma_{\mathrm{cav}} /(B^2 \Gamma_{\mathrm{cav}} +\Gamma_O)$. 
Now we see the Franck--Condon factor acting to reduce the source efficiency~\cite{Kaer2013}, 
which demonstrates a trade-off between the two source figures of merit.
Crucially, however, the increase in $\Gamma_{\mathrm{cav}}=4 g^2/\kappa_c$ with decreasing cavity width $\kappa_c$ can 
compensate for sideband photons which are lost, giving rise to an overall increase in efficiency as $\kappa_c$ is reduced. 

Considering lastly the strong coupling regime for the cavity case (iii), where $4 g>\kappa_c$, we see from Fig.~(3) that the 
indistinguishability begins to drop sharply, indicating that the cavity-based source cannot be arbitrarily 
improved by decreasing $\kappa_c$ (or increasing $g$). 
In this regime Rabi oscillations occur between the QD and cavity, allowing Markovian phonon-induced dephasing 
mechanisms to have a greater effect. Moreover, these Rabi oscillations 
give the excitation a greater probability to be lost to non-cavity modes, as seen by the corresponding drop in 
efficiency.

\section*{Discussion}

Our results allow for a critical appraisal of the most commonly used 
single photon source architectures. For a QD in a perfect lossless waveguide, although efficiencies may well 
approach $1$, even in the absence of pure-dephasing ($\gamma=0$), the broadband nature of Purcell enhancement means that 
the unavoidable phonon sideband in the emission 
spectrum limits photon indistinguishability to approximately $B^4=83\%$ at $T=4~\mathrm{K}$. 
A filter can improve this value, but the efficiency will then necessarily decrease, giving 
$\mathcal{I}\approx 99\%$ and $\eta=83\%$ for a filter width of $\hbar\kappa_f=100~\mu\mathrm{eV}$.

For a QD coupled to a cavity, we can identify an optimal regime where $4g<\kappa_c \ll \xi$, such that 
the cavity removes the sideband, but is not so narrow as to enter the strong coupling regime. Clearly 
a small QD--cavity coupling strength $g$ most easily satisfies this criterion, though this comes at the expense of 
a reduced efficiency as the cavity Purcell effect weakens. 
These competing requirements 
mean a cavity-based source cannot simultaneously reach near-unity efficiency and indistinguishability 
by simply increasing the cavity Q-factor or QD--cavity coupling strength. 
Nevertheless, readily achievable experimental values  
of $\hbar g=30~\mu\mathrm{eV}$ and $\hbar\kappa_c = 120~\mu\mathrm{eV}$ give $\mathcal{I}=99\%$ and $\eta = 96\%$ at $T=4~\mathrm{K}$.

These numbers and the calculations in Fig.~(3) are based on a favourable but realistic scenario, 
in which phonons are the dominant source of dephasing, and placing the QD in a cavity does not affect its emission into non-cavity 
modes. This immediately points us towards how source architectures may be improved, 
as the figures of merit are ultimately limited by the size of the phonon sideband in the bulk QD spectrum 
and the strength of emission into non-cavity modes. The former may be reduced in QDs 
with a larger exciton localisation length~\cite{PhysRevLett.105.177402}, 
or actively suppressed by manipulation of the phononic density of 
states. Both of these approaches, however, come at the risk of increasing ZPL dephasing~\cite{Tighineanu2017}
which must be avoided. 
Perhaps more promising is the prospect of decreasing photon emission into non-cavity modes. 
Our results suggest that future cavity designs ought to carefully take into account the 
spectrum and strength of emission into these leaky modes, as well as the 
usual cavity mode volume and Q factor. 
Decreased emission into non-cavity modes is possible for low Q-cavities~\cite{Androvitsaneas2016}, 
though these cavities will not be spectrally narrow enough to remove the sideband. 
Instead, a photonic environment that strongly suppresses all 
emission except into a spectrally narrow ($\sim 0.1~\mathrm{meV}$) cavity mode is required. 

\section*{Methods}

To find the detected and out-of-plane electric fields which determine the relevant emission 
properties we write $E_{\mu}(\omega) = \sum_l c_{\mu,l}(\omega)$, where $c_{\mu,l}(\omega)$ is 
the annihilation operator for mode $l$ of environment $\mu$ moved into the 
Heisenberg picture and Fourier transformed, 
with $\mu=\{D,O\}$ denoting the detected ($D$) and out-of-plane ($O$) channels. 
The way in which the mode operators $c_{\mu,l}$ (and hence the fields) couple to the QD 
is contained within the Hamiltonian term $H_I^{EM}$, and depends on the source architecture under consideration, 
with the full details given in the Supplementary information. In all cases, 
equations of motion coupling the electric fields to the QD degrees of freedom are 
obtained from the polaron transformed Hamiltonian, and therefore contain bath displacement operators 
which give rise to a phonon sideband. 

For case (i), a defining characteristic of slow-light waveguides is the broadband nature of the Purcell enhancement~\cite{Rao2007}. 
We therefore assume a flat photonic spectrum over frequencies relevant to the QD, 
from which we find the detected and out-of-plane fields are 
$\smash{\tilde{E}_{D,O}}(t) \approx i\smash{\sqrt{\Gamma_{D,O}/2\pi}} \tilde{\sigma}(t) \tilde{B}_-(t)$ in the time-domain, 
where $\Gamma_{D,O}$ is the corresponding emission rate, $\sigma=\ketbra{0}{X}$, and tildes indicate 
Heisenberg picture operators. 
The above expression has the same form as that of a standard quantum dipole emitter, though modified by a lattice displacement operator $B_-$, 
which through Eqs.~({\ref{EfficiencyDefinition}}) and ({\ref{IndistinguishabilityDefinition}}) affects the spectrum, 
efficiency and indistinguishability. For case (ii), the effect of a spectral filter is most easily introduced in the frequency domain,  
where the detected field becomes $E_D(\omega)= \smash{\sqrt{\Gamma_D/\Gamma_O}} h_f(\omega) E_O(\omega)$~\cite{Eberly1977}, and for a Lorentzian filter we 
have $h_f(\omega) = (\kappa_f/2)[i(\omega-\omega_f)-(\kappa_f/2)]^{-1}$ with $\kappa_f$ and $\omega_f$ the filter width and central frequency 
respectively. Introducing the filter in this way requires that we add a term 
$(\Gamma_D/\Gamma_O)\smash{\int_{-\infty}^{\infty}d\omega [1-|h_f(\omega)|^2]S_O(\omega,\omega)}$ to 
the denominator in Eq.~({\ref{EfficiencyDefinition}}) to include the field rejected by the filter. 
In the time domain the detected field takes the form of a convolution between the emitted field and the filter response function.

We follow a similar procedure 
for case (iii), though now explicitly account for variation of the cavity lineshape across the relevant QD frequencies. 
The out-of-plane emission (i.e. not via the cavity mode) is given by 
$\tilde{E}_{O}(t) \approx i\sqrt{\Gamma_{O}/2\pi} \tilde{\sigma}(t) \tilde{B}_-(t)$, 
which takes the same form as in case (i). 
We make the assumption that the detected field consists of 
those photons emitted by the cavity mode~\cite{Grange2015,Grange2016,Unsleber2015,PhysRevB.87.081308}. 
Although it is customary to define the detected field in this way for QD--cavity systems, 
one expects that in the very broad cavity limit the detected field will also contain a contribution arising from 
direct QD emission. We do not include this contribution in our calculations, though note that 
its effect would be to slightly raise efficiencies in the less interesting $\kappa_c\gg 4 g$ regime for case (iii). 
Taking the usual detected field definition, we find it can be written in frequency space as  
$E_{D} (\omega)  = \smash{\sqrt{4g^2/\kappa_c\Gamma_O}}h_c(\omega) E_{O} (\omega)$, with 
$h_c(\omega) = i(\kappa_c/2)[i(\omega - \omega_{c} )-\kappa_c/2]^{-1}$, where $g$ is the light--matter coupling strength, 
$\kappa_c$ the cavity width, and $\omega_c$ the cavity mode frequency. 
Comparing to case (ii) above, 
this expression demonstrates the analogy between a cavity and a spectral filter, 
and the mathematical connection between filtering effects 
and the phonon sideband captured in the operator $\tilde{B}_-(t)$. One can see that coupling to a cavity has two 
dominant effects. The first is to modify the QD dynamics, which is captured implicitly in 
the time-dependence of the operator $\tilde{\sigma}(t)$. How these dynamics are modified will depend on the regime 
of light--matter coupling, and will include Purcell enhancement, as well as 
phonon induced dephasing mechanisms~\cite{PhysRevB.87.081308}. 
The second is to spectrally filter the resulting QD emission, as described 
by the cavity filter function $h_c(\omega)$. 
With these relationships between the electric fields, 
it follows that the spectra can be written $S_D(\omega,\nu)=\mathcal{G}(\omega,\nu)S_O(\omega,\nu)$. 

Finally, we note that the relationship $S_D(\omega,\nu)=\mathcal{G}(\omega,\nu)S_O(\omega,\nu)$ 
is exact in cases (i) and (ii). In case (iii) it is exact in the absence of coupling to phonons, valid 
in both the strong and weak QD--cavity coupling regimes. As discussed in detail in the supplementary 
information, when phonons are included, the theory remains quantitatively accurate 
except in the very strong coupling regime where dissipative terms in the master equation 
not included in the Green's function $\mathcal{G}(\omega,\nu)$ become important. Nevertheless, 
in this regime the present theory remains qualitatively accurate when compared to an exact approach, 
and correctly predicts the fall in source merit criteria with decreasing cavity width.

\subsection*{Data availability}

The data that support the plots within this paper and other findings of this study 
are available from the corresponding author upon reasonable request.

%

\providecommand{\noopsort}[1]{}\providecommand{\singleletter}[1]{#1}%

\section*{Acknowledgements}

The authors wish to thank Niels Gregersen for useful discussions. 
J.I.-S. and J.M. acknowledge support from the Danish Research Council (DFF-4181-00416) and 
Villum Fonden (NATEC Centre). A.N. is supported by University of Manchester and the Engineering and 
Physical Sciences Research Council, grant number EP/N008154/1. This project has received funding from the 
European Union's Horizon 2020 research and innovation programme under the 
Marie Sk{\l}odowska-Curie grant agreement No. 703193.\\

\section*{Author contributions}

J.I.-S. and D.P.S.M. contributed equally to this work. All authors 
were involved in the conception, development, and writing of the manuscript. 

\section*{Additional information}

Supplementary material is available for this work. Correspondence and requests 
for materials should be addressed to J.I.-S.

\section*{Competing financial interests}

The authors declare no competing financial interests.

\clearpage

\widetext
\section*{\large Supplemental Material}

We consider a quantum dot (QD) with ground state $\ket{0}$ and excited (single exciton) state $\ket{X}$ 
with energy $\hbar\omega_X$. The QD couples to a phonon and photon environment, and is described by the Hamiltonian
\begin{equation}\label{ham}
H = \hbar \omega_X \ket{X}\bra{X} + \hbar\ket{X}\bra{X}\sum_k g_k(b_k^\dagger +b_k) + H^{EM}_I+H_E^{EM}+H_E^{PH},
\end{equation}
with $\smash{b_k^\dagger}$ ($b_k$) the creation (annihilation) operator 
for a phonon with wavevector $k$, and couples to the QD through the coupling constant $g_k$. 
The phonon environment is described by $H_E^{PH} = \hbar\sum_k\nu_k b^\dagger_k b_k$. The photonic (electromagnetic) 
environment is described by $H_E^{EM}$, and couples to the QD via the interaction term $H^{EM}_I$. 

\section{QD coupled to a photonic waveguide}

We first consider a QD coupled to a photonic waveguide, for which we have 
$H_{E}^{EM} = \hbar\sum_{\mu}\sum_{l}\omega_{\mu,l} c^\dagger_{\mu,l}c_{\mu,l}$, and 
\begin{equation}
H^{EM}_I= \hbar\sum\limits_{\mu }\sum\limits_l \left(f_{\mu,l}\sigma c_{\mu,l}^\dagger + f_{\mu,l}^\ast\sigma^\dagger c_{\mu,l} \right),
\end{equation}
where $\sigma = \ket{0}\bra{X}$, $c_{\mu,l}^\dagger$ ($c_{\mu,l}$) is the creation (annihilation) operator for mode $l$ of environment $\mu$ with frequency $\omega_{\mu,l}$, 
with $\mu = \{D,O\}$ denoting detected ($D$) waveguide modes, and out-of-plane ($O$) modes leading to loss. 
We characterise the QD--photon coupling strength with the spectral density $J_\mu(\omega) =\sum_l \vert f_{\mu,l}\vert^2\delta(\omega-\omega_{\mu,l})$, 
which is taken to be flat over frequency scales relevant to the QD~\cite{Rao2007}, for both the out-of-plane field and detected waveguide modes, 
allowing us to write $J_{\mu}(\omega)\approx \Gamma_\mu/\pi$, where $\Gamma_\mu$ is the emission rate into the relevant channel.

We apply the polaron transformation to $H$, defined through $H_P=\mathcal{U}H\mathcal{U}^{\dagger}$, with 
$\mathcal{U} = \outter{0} + \outter{X}B_{+}$, where $B_\pm = \exp\left(\pm\sum_k g_k(b_k^\dagger - b_k)/\nu_k\right)$, which allows us 
to derive a master equation that is non-perturbative in the electron-phonon coupling strength~\cite{nazir2015modelling}. 
The transformed Hamiltonian reads $H_P=\hbar\tilde\omega_X \ket{X}\bra{X} + H_{IP} + H_E^{EM}+H_E^{PH}$ with 
\begin{equation}
H_{IP}=\hbar\sum\limits_{\mu }\sum\limits_l \left(f_{\mu,l}\sigma c_{\mu,l}^\dagger  B_- + f_{\mu,l}^\ast\sigma^\dagger c_{\mu,l}B_+ \right),
\end{equation}
and $\tilde\omega_X = \omega_X + \sum_k g_k^2/\nu_k$, and we note the presence of phonon displacement operators in this exciton--photon coupling term. 
Physically, these operators lead to a displacement of the phonon lattice as discussed in the main text. 

We now derive a Born--Markov master equation in the polaron frame, treating $H_{IP}$ to second order 
and assuming the electromagnetic environments remain in the vacuum state. 
After moving into a rotating frame this becomes~\cite{breuer2007theory} 
\begin{equation}
\label{WGMEQ}  
\frac{\partial\rho(t)}{\partial t} = 
\Gamma_{\mathrm{tot}}\mathcal{L}_\sigma[\rho(t)] + 2\gamma\mathcal{L}_{\sigma^\dagger\sigma}[\rho(t)],
\end{equation}
where $\rho(t)$ denotes the QD density operator, $\mathcal{L}_A[\rho]=A\rho A^{\dagger}-(1/2)(A^{\dagger}A\rho + \rho A^{\dagger}A)$,  
the total emission rate is $\Gamma_{\mathrm{tot}} = \Gamma_O + \Gamma_D$, and we 
have introduced a phenomenological dephasing rate $\gamma$ to capture broadening of the zero-phonon-line not caused by phonons.

\subsection*{Field emission for cases (i) and (ii)}

As defined in Eqs. (1) and (2) in the main text, the indistinguishability and efficiency are calculated 
from the detected and out-of-plane two colour spectra, $S_{\mu}(\omega,\nu)=\langle E_{\mu}(\omega)^{\dagger}E_{\mu}(\nu)\rangle$, which are 
Fourier transforms of the first order two-time correlation function 
$g_\mu^{(1)}(t_1,t_2) = \langle \tilde{E}_\mu^\dagger(t_1)\tilde{E}_\mu(t_2)\rangle$, 
where $\tilde{E}_\mu(t) = \sum_{l} \tilde{c}_{\mu,l}(t)$ is the electric field operator in the Heisenberg picture. 
The Heisenberg equations of motion give 
\begin{equation}
\dot{\tilde{c}}_{\mu,l}(t) = -i\omega_{\mu,l} \tilde{c}_{\mu,l}(t) - i f_{\mu,l} \tilde{\sigma}(t) \tilde{B}_-(t), 
\end{equation}
which can be formally solved to give $\tilde{E}_\mu(t) = \sqrt{\frac{\Gamma_\mu}{2\pi}}\tilde{\sigma}(t) \tilde{B}_-(t)$, 
where we have and neglected the vacuum contribution. We can therefore write 
$S_{D}(\omega,\nu)=(\Gamma_D/\Gamma_{O})S_{O}(\omega,\nu)$ leaving us to calculate 
$S_{O}(\omega,\nu)$ as described in Section~{\ref{IndAndEffSection}}. 

Considering now the filtered waveguide source, the effect of the filter on the detected field is introduced by letting 
$E_D(\omega) = \sqrt{\Gamma_D/\Gamma_O}h_f(\omega)E_O(\omega)$~\cite{Eberly1977}, where for a Lorentzian filter we have  
$h_f(\omega)=(\kappa_f/2)[(\kappa_f/2)-i(\omega-\omega_0)]^{-1}$, with $\omega_f$ and $\kappa_f$ the filter central frequency 
and width, respectively. This allows us to write the detected two-colour spectrum as 
$S_{D}(\omega,\nu)=(\Gamma_D/\Gamma_{O}) h_f^*(\omega) h_f(\nu) S_{O}(\omega,\nu)$.

\section{QD coupled to an optical cavity}

We now consider the QD coupled to a one-sided single mode cavity with fundamental frequency $\omega_c$ and width $\kappa_c$, defined by creation (annihilation) 
operator $a^\dagger$ ($a$). The QD also couples to non-cavity modes (again labeled with and $O$ subscript) giving rise to loss, while the cavity couples to detected modes (labeled 
with a $D$ subscript). In the rotating wave approximation the exciton--photon interaction Hamiltonian now takes the form
\begin{equation}
H^{EM}_I=\hbar g\left(\sigma^\dagger a + \sigma a^\dagger\right)
+ \hbar\sum\limits_l \left(f_{O,l}\sigma^\dagger c_{O,l} + \text{h.c.}\right) +  \hbar\sum\limits_l \left(f_{D,l}a^\dagger c_{D,l} + \text{h.c.}\right)
\end{equation}
where $\text{h.c.}$ denotes the Hermitian conjugate, $g$ is the QD--cavity coupling strength, 
and $f_{\mu,l}$ is the coupling strength between the QD ($\mu=O$) or cavity ($\mu=D$) and mode $l$ of the relevant environment.
The energy of the environments and cavity is described by $H_E^{EM}= \hbar\omega_c a^\dagger a + \hbar\sum_{\mu} \sum_l  \omega_{\mu,l} c^\dagger_{\mu,l} c_{\mu,l}$. 
As for the waveguide case we assume that lossy and detected modes have a flat spectral density, such that 
$J_O(\omega)\approx  \Gamma_{O}/\pi$, and $J_D(\omega)\approx \kappa_c/\pi$.

Applying the same polaron transformation to the full QD--cavity Hamiltonian, we find~\cite{Iles-Smith:16}
\begin{equation}
H_P =\hbar\tilde\omega_X\outter{X} + \hbar g_r\hat{X} + H_E+ \hbar g(\hat{X} B_X + \hat{Y}B_Y) 
+\hbar\sum_l \left(f_{O,l}\sigma^\dagger c_{O,l}B_+ + \text{h.c.}\right) +  \hbar\sum_l \left(f_{D,l}a^\dagger c_{D,l} + \text{h.c.}\right)
\end{equation}
where we have defined the operators $\hat{X} = \sigma^\dagger a+ a^{\dagger}\sigma$, 
$\hat{Y} = i(\sigma^\dagger a - \sigma a^\dagger)$, $B_X = (B_+ + B_- - 2B)/2$, and $B_Y = i(B_+-B_-)/2$, 
and $g_r = gB$ with $B=\langle B_{\pm}\rangle$ is the renormalised QD--cavity coupling strength. 
We now derive a Born-Markov master equation in the polaron frame, treating the last three terms in $H_P$ to second order and 
assuming the phonons to be in a thermal state described by 
$\rho^{PH}_E = \exp(-\hbar\beta\sum_k\nu_kb_k^\dagger b_k)/\tr_E[\exp(-\hbar\beta\sum_k\nu_kb_k^\dagger b_k)]$ at inverse temperature $\beta = 1/(k_B T)$.
Taking the cavity to be on resonance with the polaron shifted QD transition energy, $\omega_c= \tilde{\omega}_X$, 
and in a frame rotating at this frequency, we find
\begin{equation}
\label{CavityME}
\frac{\partial \rho(t)}{\partial t} = -i [ g_r \hat{X},\rho(t)]
+\mathcal{K}_{\mathrm{ph}}[\rho(t)]+ \kappa_c\mathcal{L}_{a}[\rho(t)] 
+ \Gamma_O\mathcal{L}_{\sigma}[\rho(t)]+2\gamma\mathcal{L}_{\sigma^\dagger\sigma}[\rho(t)]
\end{equation}
where $\rho(t)$ is the reduced density operator describing the QD and cavity mode, and 
we have again introduced a phenomenological pure dephasing term $\gamma$. 
The phonon dissipator may be written as
\begin{equation}
\mathcal{K}_{\mathrm{ph}} = - g^2\left(\left[\hat{X},\hat{X}\rho(t)\right]  \chi_X+ \left[\hat{Y},\hat{Y}\rho(t)\right]  \chi_Y+\left[\hat{Y},\hat{Z}\rho(t)\right] \chi_Z
+\text{h.c.}\right) 
\end{equation}
with $\hat{Z} = \sigma^\dagger\sigma-a^\dagger a$. 
Here we have defined $\chi_{X} =\int_0^\infty d\tau \Lambda_{XX}(\tau)$, $\chi_{Y} =\int_0^\infty d\tau\cos(2g_r\tau) \Lambda_{YY}(\tau)$, 
and $\chi_{Z} =-\int_0^\infty d\tau\sin(2g_r\tau) \Lambda_{YY}(\tau)$, with phonon correlation functions defined as 
$\Lambda_{XX} (t) = \langle B_X(\tau)B_X\rangle = B^2(e^{\varphi(t)} + e^{-\varphi(t)}-2)/2$ and 
$\Lambda_{YY} (t) = \langle B_Y(\tau)B_Y\rangle = B^2(e^{\varphi(t)} - e^{-\varphi(t)})/2$, where 
\begin{equation}
\varphi(t)=\int_0^{\infty}d\omega J_{\mathrm{ph}}(\omega)\omega^{-2}[\coth(\hbar \beta\omega/2)\cos(\omega t)-i\sin(\omega t)].
\label{propagator}
\end{equation}
Note that the above master equation is non-perturbative in the electron-phonon coupling strength, 
but relies on the light matter coupling strength being small compared to the environmental cut-off frequency, 
that is $g/\xi\ll1$, which is satisfied for realistic QD cavity systems.

\subsection*{Field emission in case (iii)}

Following a similar procedure as in the waveguide case, we solve the Heisenberg equation of motion for 
the emitted field operators in terms of the operators pertaining to the QD and cavity.
Emission directly from the QD leads to a field operator of the form $\tilde{E}_{O}(t)=\sqrt{\Gamma_O/2\pi} \tilde{\sigma}(t)\tilde{B}_-(t)$, 
which has the same form as the waveguide emission, with the displacement operators leading to the phonon sideband.


To write the detected field in terms of QD emission, we solve the Heisenberg equation of motion for the cavity operator in frequency space. 
The Heisenberg equations of motion give
\begin{equation}
\frac{\partial \tilde{a}(t)}{\partial t} = -i\omega_c \tilde{a}(t) - ig \tilde{\sigma}(t) \tilde{B}_-(t) -i\sum\limits_l f_{D,l} \tilde{c}_{D,l}(t),
\qquad
\mathrm{and}
\qquad
\frac{\partial \tilde{c}_{D,l}(t)}{\partial t} =-i\omega_{D,l} \tilde{c}_{D,l}(t) - if_{D,l}\tilde{a}(t).
\end{equation}
Moving into the Fourier domain and solving algebraically we find 
\begin{equation}
{c}_{D,l}(z) = \frac{f_{D,l}}{z-\omega_{D,l}}{a}(z),
\qquad
\mathrm{and}
\qquad
{a}(z) = \frac{ig\zeta(z)}{i(z-\omega_{c}) - \kappa_c/2},
\end{equation}
where ${O} (z)= \int_{-\infty}^{\infty} dt e^{izt}\tilde{O}(t)$ with $z$ is the Fourier space variable, 
and ${\zeta}(z) = \int_{-\infty}^{\infty} dt e^{izt}\tilde{\sigma}(t)\tilde{B}_-(t)$. 
We have used the Kramers-Kronig relations to write 
$\sum_l\vert f_{D,l}\vert^2/(\omega_{D,l}-z)\rightarrow\int_{-\infty}^\infty d\omega J_{D}(\omega)/(\omega-z) = i\pi J_{D}(z)\approx i\kappa_c/2$.
The detected field can now be written $E_D(z) =\sqrt{4g^2/\kappa_c\Gamma_O} h_c(z)E_{O}(z)$, 
where $h_c(z) = i(\kappa_c/2)[i(z-\omega_c) - \kappa_c/2]^{-1}$ and $E_{O}(z) = \sqrt{\Gamma_O/2\pi}\zeta(z)$.
Notice the phonon displacement operators are now present in the expression of the detected field, 
and the cavity acts as a filter of the QD emission.
As in the filtered waveguide case, the indistinguishability may be calculated directly from the 
detected generalised two colour spectrum, which for the cavity takes the form 
$\mathcal{S}_D(\omega,\nu) = (\Gamma_{\mathrm{cav}}/\Gamma_O)h_c^\ast(\omega)h_c(\nu)S_O(\omega,\nu)$, 
with $\Gamma_{\mathrm{cav}}=4 g^2/\kappa_c$.

We note that writing the cavity spectrum in terms of the QD spectrum in this way is not restricted 
to the weak exciton--photon coupling regime. In fact, in the absence of coupling to phonons, 
the relation 
$\langle a^{\dagger}(\omega)a(\nu)\rangle = h_c^\ast(\omega)h_c(\nu) \langle \sigma^{\dagger}(\omega)\sigma(\nu)\rangle$ 
can be shown to be exact given our initial condition consisting of one excitation in the QD, since 
in this case the two spectra can be calculated exactly using the regression theorem. 
The inclusion of phonons introduces extra dissipative terms in the master equation which are not adequately captured 
when solving the Heisenberg equations of motion as described above, meaning the 
relations $\langle a^{\dagger}(\omega)a(\nu)\rangle = h_c^\ast(\omega)h_c(\nu) \langle \sigma^{\dagger}(\omega)\sigma(\nu)\rangle$ and 
therefore $\mathcal{S}_D(\omega,\nu) = (\Gamma_{\mathrm{cav}}/\Gamma_O)h_c^\ast(\omega)h_c(\nu)S_O(\omega,\nu)$ become 
approximate. With this in mind, we can expect our formalism as presented to be accurate 
accept when $\mathcal{K}_{\mathrm{ph}}$ in Eq.~({\ref{CavityME}}) becomes dominant. 
This is discussed in more detail below. 

\section{Indistinguishability and efficiencies}
\label{IndAndEffSection}

In order to calculate the detected indistinguishability and efficiency defined in Eqs.~(1) and (2) in the main text, we must calculate the detected and lost two-colour spectra 
$S_{D,O}(\omega,\nu)=\langle E_{D,O}^{\dagger}(\omega) E_{D,O}(\nu)\rangle$. In all cases we can write $S_{D}(\omega,\nu)=\mathcal{G}(\omega,\nu)S_{O}(\omega,\nu)$, 
with $\mathcal{G}(\omega,\nu)=(\Gamma_D/\Gamma_O)$ in case (i) for the waveguide, $\mathcal{G}(\omega,\nu)=(\Gamma_D/\Gamma_O)h_f^*(\omega) h_f(\nu)$ for case 
(ii) for the filtered waveguide source, and $\mathcal{G}(\omega,\nu)=(\Gamma_{\mathrm{cav}}/\Gamma_O)h_c^*(\omega) h_c(\nu)$ for case (iii) with the cavity. 

It therefore suffices to calculate the spectrum
$S_O(\omega,\nu)=\int_{-\infty}^{\infty} d t_1 \int_{-\infty}^{\infty} d t_2 \langle \tilde{E}_O^{\dagger} (t_1) \tilde{E}_O (t_2)\rangle \mathrm{e}^{- i \omega t_1}\mathrm{e}^{i \omega t_2}$.
In all cases we have found $\tilde{E}_O(t)=\sqrt{  \Gamma_O/2\pi} \tilde{\sigma}(t) \tilde{B}_-(t)$, and we therefore define the first order correlation function 
\begin{equation}
g^{(1)}_O(t_1,t_2) =\frac{ \Gamma_O}{2\pi}\left\langle \tilde{\sigma}^\dagger(t_1)\tilde{B}_+(t_1)\tilde{\sigma}(t_2)\tilde{B}_-(t_2)\right\rangle. 
\label{Factorisedg1}
\end{equation} 
The many body displacement operators in this expression make it challenging to calculate exactly. 
However, for QDs the timescales associated with phonon relaxation and photon emission are very different, with the 
former typically occurring on a picosecond timescale, while excitonic recombination occurs on a $100~\mathrm{ps}$ to $1~\mathrm{ns}$ timescale. 
This allows us to factorise the correlation function to give 
\begin{equation}
g^{(1)}_O(t_1,t_2)\approx \frac{ \Gamma_O}{2\pi}G^{\ast}(t_1,t_2)g^{(1)}(t_1,t_2). 
\end{equation}
The factor $G(t_1,t_2)=\left\langle B_+(t_1)B_-(t_2)\right\rangle$ is the two-time correlation function of the phonon environment, while 
$g^{(1)}(t_1,t_2)=\left\langle \sigma^\dagger(t_1)\sigma(t_2)\right\rangle$ is the two-time correlation function for a QD dipole. 
Using the standard procedure we find $G(t+\tau,t)\equiv G(\tau)=B^2 \exp[\varphi(\tau)]$ with $\varphi(\tau)$ already 
defined in Eq.~({\ref{propagator}}), and where $B^2=\exp[-\varphi(0)]$ is the Frank--Condon factor. 

We now make use of the differing timescales of the optical and vibrational processes to separate out the phonon and dipole contributions to 
the bare spectrum ${S}_O(\omega,\nu) = {S}_{\mathrm{ZPL}}(\omega,\nu) +  {S}_{\mathrm{SB}}(\omega,\nu)$, 
where ${S}_{\mathrm{ZPL}}(\omega,\nu) = \mathcal{S}_{\mathrm{ZPL}}(\omega,\nu) + \mathcal{S}_{\mathrm{ZPL}}(\nu,\omega)^\ast$ 
and ${S}_{SB}(\omega,\nu) = \mathcal{S}_{\mathrm{SB}}(\omega,\nu) + \mathcal{S}_{\mathrm{SB}}(\nu,\omega)^\ast$, with
\begin{equation}
\begin{split}
&\mathcal{S}_{\mathrm{ZPL}}(\omega,\nu)= B^2\Gamma_O \int_0^\infty dt
\int_0^\infty d\tau
\mathrm{e}^{i(\nu-\omega) t}\mathrm{e}^{-i\omega \tau}g^{(1)}(t+\tau,t),\\
&\mathcal{S}_{\mathrm{SB}}(\omega,\nu) = \Gamma_O\int_0^\infty dt \mathrm{e}^{i(\nu-\omega) t}g^{(1)}(t,t)\int_0^\infty d\tau (G(\tau)-B^2)\mathrm{e}^{-i\omega\tau}.
\end{split}
\end{equation}
This separation of the phonon sideband and the zero-phonon line is valid when the dynamical time-scales associated with the ZPL are much slower than the phonon relaxation. 
The final ingredient necessary is the correlation function $g^{(1)}(t_1,t_2)=\left\langle \sigma^\dagger(t_1)\sigma(t_2)\right\rangle$, which we calculate 
with use of the quantum regression theorem and Eq.~({\ref{WGMEQ}}) for cases (i) and (ii) with the waveguide, and Eq.~({\ref{CavityME}}) for case (iii) with the cavity.
The expressions quoted above are those used to plot the indistinguishability and efficiencies in Fig.~(3) in the main text. 

\begin{figure}
	\includegraphics[width =0.75\textwidth]{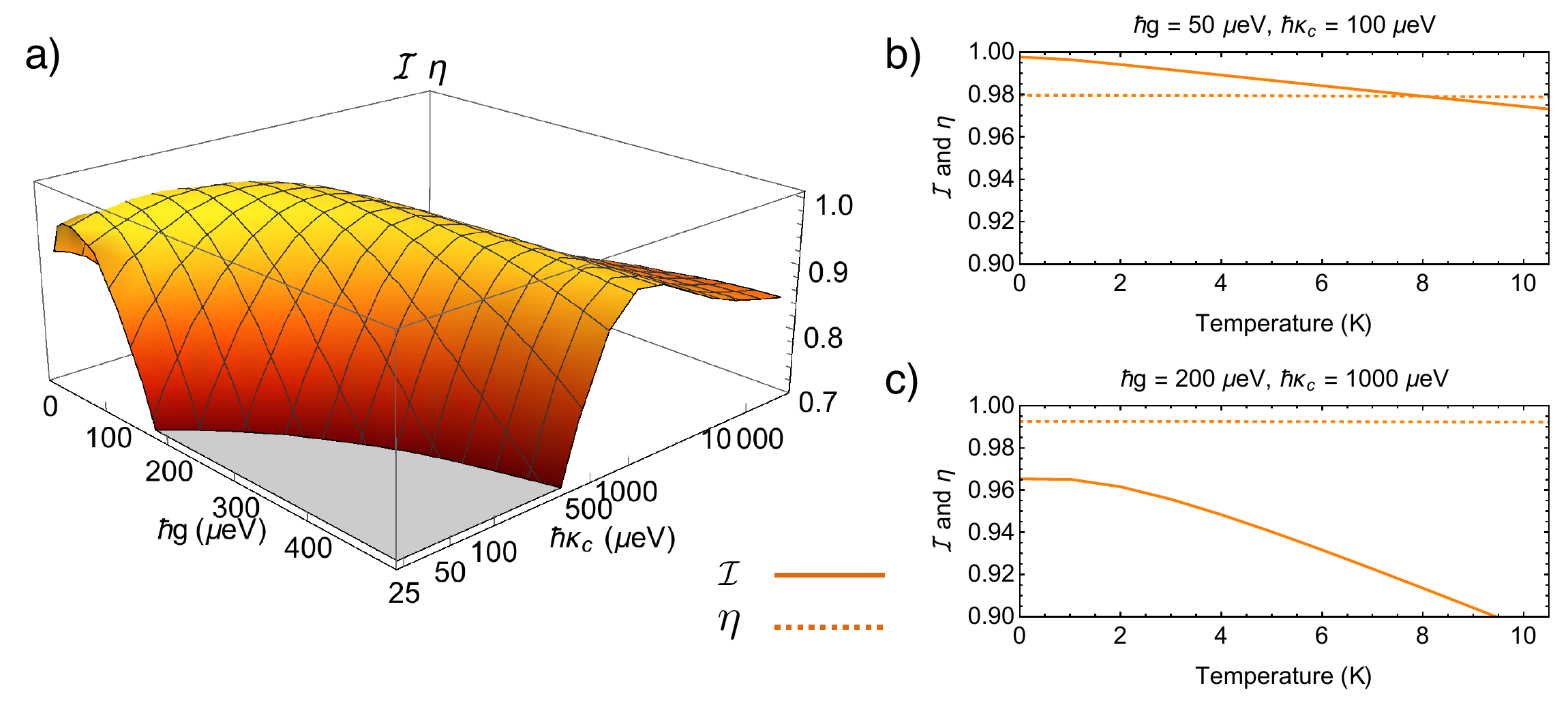}
	\caption{Part a) shows the product of the indistinguishability $\mathcal{I}$ and efficiency $\eta$ as a function of 
		the QD--cavity coupling strength $g$ and cavity decay rate $\kappa_c$ at $T=0~\mathrm{K}$. The maximum 
		occurs for approximately $\hbar g=50~\mu\mathrm{eV}$ and $\hbar\kappa_c=100~\mu\mathrm{eV}$, where 
		we then plot $\mathcal{I}$ (solid) and $\eta$ (dotted) as a function of temperature in part b). Part c) 
		shows the merit criteria as a function of temperature for a different parameter set as indicated. 
		Other parameters as in the main text, $\hbar\Gamma_O = 1~\mu\mathrm{eV}$, 
		$\alpha = 0.03~\mathrm{ps}^2$ and $\hbar \xi = 1.45~\mathrm{meV}$.}
	\label{ParamSweep}
\end{figure}

In Fig.~{\ref{ParamSweep}}~(a) of this supplement we use these expressions to calculate the product of the indistinguishability and 
efficiency across a broad range of QD--cavity coupling strengths and cavity decay rates at $T=0~\mathrm{K}$. 
In parts (b) and (c) we show how the merit criteria behave as a function of temperature for two parameter sets showing 
high merit criteria. As also discussed in the main text, the regime in which both indistinguishability 
and efficiency are high appears to be away from the strong coupling regime (away from $4g>\kappa_c$). 
Using this initial parameter search we focus on representative values of $\hbar g = 50,100,200~\mu\mathrm{eV}$ 
and a realistic temperature of $T=4~\mathrm{K}$ in the main text. 

\begin{figure}[b]
	\includegraphics[width =0.75\textwidth]{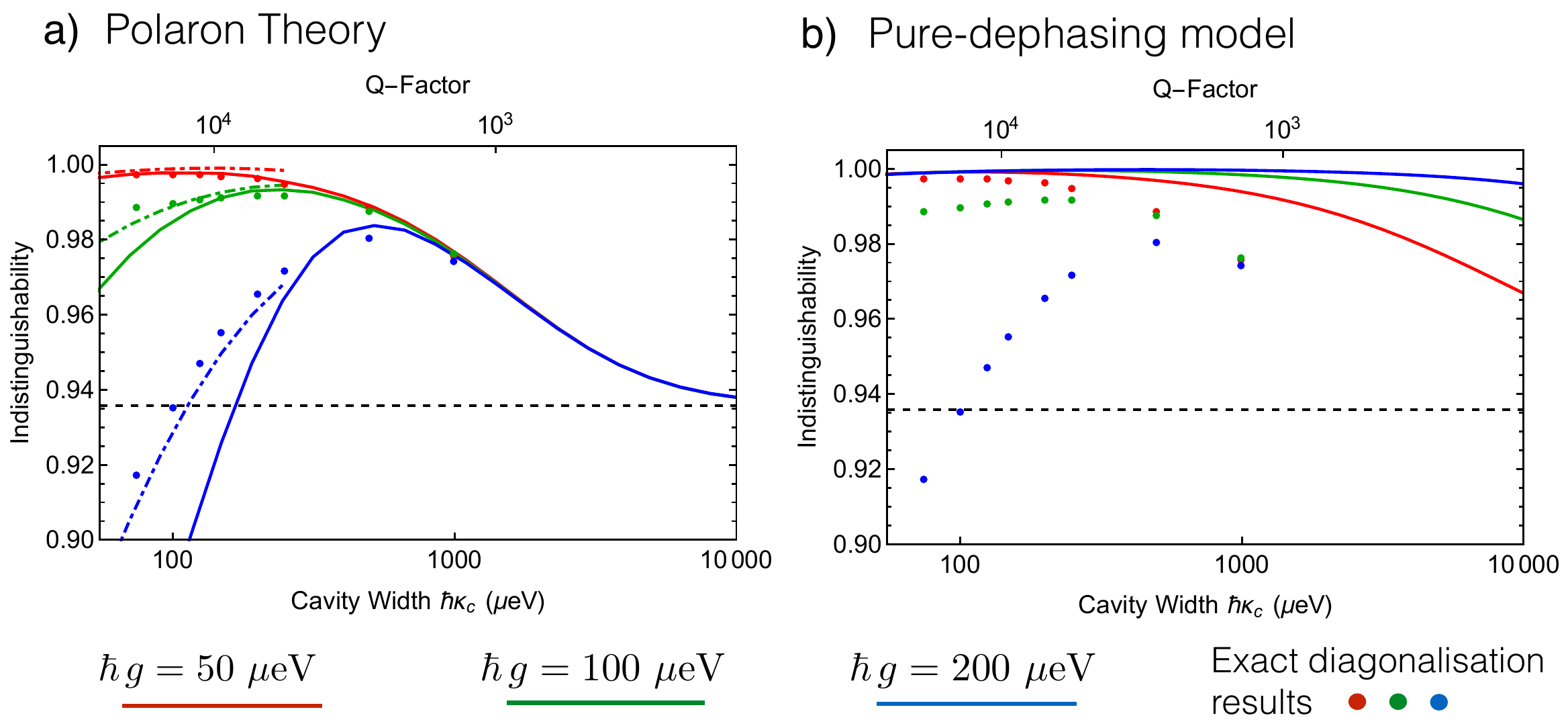}
	\caption{Part a) compares the present polaron theory (solid curves) to the exact diagonalisation results of 
		Ref.~\cite{PhysRevB.90.035312} (plot markers), while the dashed black curve shows the analytic result 
		valid in the $\kappa_c\to\infty$ limit. In the strong coupling limit where $\kappa_c< 4g$ the agreement becomes 
		only qualitative, though we note that this can be corrected with a modified polaron theory shown 
		by the dot-dashed curves. Part b) shows the phenomenological pure-dephasing theory, demonstrating qualitatively incorrect 
		behaviour in all regimes. The parameters used are $\hbar\Gamma_O = 1~\mu\mathrm{eV}$, 
		$\alpha = 0.032~\mathrm{ps}^2$, $\hbar \xi = 0.95~\mathrm{meV}$, and $T=0~\mathrm{K}$}
	\label{TheoryComparison}
\end{figure}

In order to assess the validity of our polaron theory, in Fig.~(\ref{TheoryComparison}-a) we compare it (solid curves) 
to the exact diagonalisation results of Ref.~{\cite{PhysRevB.90.035312}} (plot markers), while also showing the 
exact analytic result valid in the large $\kappa_c$ limit with the black dashed line. We note that the polaron theory is 
quantitatively accurate in all regimes except when both the QD--cavity coupling $g$ is large, and the cavity decay 
rate $\kappa_c$ is small. We emphasise that in this very strong coupling regime the source indistinguishability is 
not high according to the exact diagonalisation calculations, and the polaron theory correctly captures this behaviour 
qualitatively. Furthermore, if necessary our theory can be easily corrected in this regime by using 
the cavity operators directly for the detected field spectrum, i.e. using 
$S_D(\omega,\nu)\propto \langle a^{\dagger}(\omega) a(\nu)\rangle $ rather 
than $S_D(\omega,\nu)=\mathcal{G}(\omega,\nu)S_O(\omega,\nu)$ and Eq.~({\ref{Factorisedg1}}), as shown by the dot-dashed curves. 
We opt not to use this corrected theory in the main text as the analogy to the filter is weakened, and using 
Markovian regression leads to inaccurate results in the more relevant regime of interest where $4 g <\kappa_c$. 
Finally, we note that our results suggest the most interesting regime is when the QD--cavity coupling strength 
is small to moderate, where our theory is quantitatively accurate. This is 
supported by those experiments claiming the highest indistinguishability, most 
notably Ref.~\cite{Somaschi2016} which has  
$\hbar g \approx 10~\mu\mathrm{eV}$ and $\hbar\kappa_c\approx 120~\mu\mathrm{eV}$, while 
Ref.~\cite{Ding2016} has $\hbar g \approx 20~\mu\mathrm{eV}$ and $\hbar\kappa_c\approx 230~\mu\mathrm{eV}$

By means of comparison in Fig.~({\ref{TheoryComparison}}-b) we consider the accuracy a phenomenological pure-dephasing theory, 
obtained by setting the phonon terms $\mathcal{K}_{\mathrm{ph}}=0$ in Eq.~({\ref{CavityME}}), and choosing  
the pure-dephasing rate $\gamma$ such that in the very broad cavity limit the same indistinguishability value is 
found as in the full phonon theories. As $\kappa_c$ is reduced from large values, 
in contrast to the more sophisticated theories for which cavity filtration effects dominate, in the pure-dephasing theory 
Purcell enhancement is the most important effect. As such, the pure-dephasing theory incorrectly predicts 
that larger values of $g$ are preferred. It also incorrectly predicts how narrow the cavity must be, 
reflecting that in this theory a reduced cavity width only increases the Purcell effect, rather than increasing 
cavity filtering properties which are important when the sideband is included.

\subsection*{Analytic approximations in the adiabatic limit}

In order to obtain some analytic expressions, we consider the regime 
that the filter or cavity are broad enough that features in the vicinity of the ZPL are unaffected. This corresponds 
to $\Gamma_D<\kappa_f$ for the filter, and $\Gamma_{\mathrm{cav}}< \kappa_c$ for the cavity. 
In the cavity case, when $\kappa_c\gg g$, the cavity may be adiabatically eliminated, which is to assume it always remains in its steady state. 
Thus, we may replace the cavity operator with its steady state value $a \approx 2ig_r\sigma/\kappa$. 
In this case the cavity-QD operators become $\hat{X} = \sigma^\dagger a +\text{h.c.}\approx 0$, 
$\hat{Y} = i(\sigma^\dagger a -\text{h.c})\approx 4\kappa_c^{-1}g_r\sigma^\dagger\sigma$, and 
$\hat{Z} = \sigma^\dagger\sigma-a^\dagger a\approx(1-4\kappa^{-2}g_r^2)\sigma^\dagger\sigma$. 
If we use these expressions in Eq.~({\ref{CavityME}}), we obtain
\begin{equation}
\frac{\partial\rho(t)}{\partial t} \approx
\Gamma_{\mathrm{tot}}\mcL_{\sigma}[\rho(t)] + 2\gamma_{\mathrm{tot}}\mcL_{\sigma^\dagger\sigma}[\rho(t)]
\label{CavMEAd}
\end{equation} 
where $\rho(t)$ is now the reduced state of the QD, $\Gamma_{\mathrm{tot}}=\Gamma_O + \Gamma_{\mathrm{cav}}$ with 
$\Gamma_{\mathrm{cav}}=4 g^2/\kappa_c$ is the total emission rate, and $\gamma_{\mathrm{tot}}=\gamma_{\mathrm{ph}}+\gamma$, 
with $\gamma_{\mathrm{ph}}=(4g_r/\kappa_c)^2\gamma_Y
+ \gamma_Z(4g_r/\kappa_c)\left(1-(2g_r/\kappa_c)^2\right)$, while $\gamma_Y=\mathrm{Re}[\chi_Y]$ and $\gamma_Z=\mathrm{Re}[\chi_Z]$. 
The term proportional to $\gamma_Z$ is negligible in the regime $\hbar g<k_BT<\hbar\xi$~\cite{McCutcheon2013}, and 
we may then write $\gamma_{\mathrm{ph}}\approx(4g_r/\kappa_c)^2\gamma_Y$.
This expression can be further simplified in the limit of weak electron-phonon coupling, where $\exp\left(\varphi(\tau)\right)\approx 1-\varphi(\tau)$. 
Performing the integration over $\tau$ we obtain $\gamma_{\mathrm{ph}} \approx 2\pi\left(gB/\kappa_c\right)^2J_{\mathrm{ph}}(2gB)\coth\left(\hbar gB/k_BT\right)$.

\begin{figure}
	\includegraphics[width =0.75\textwidth]{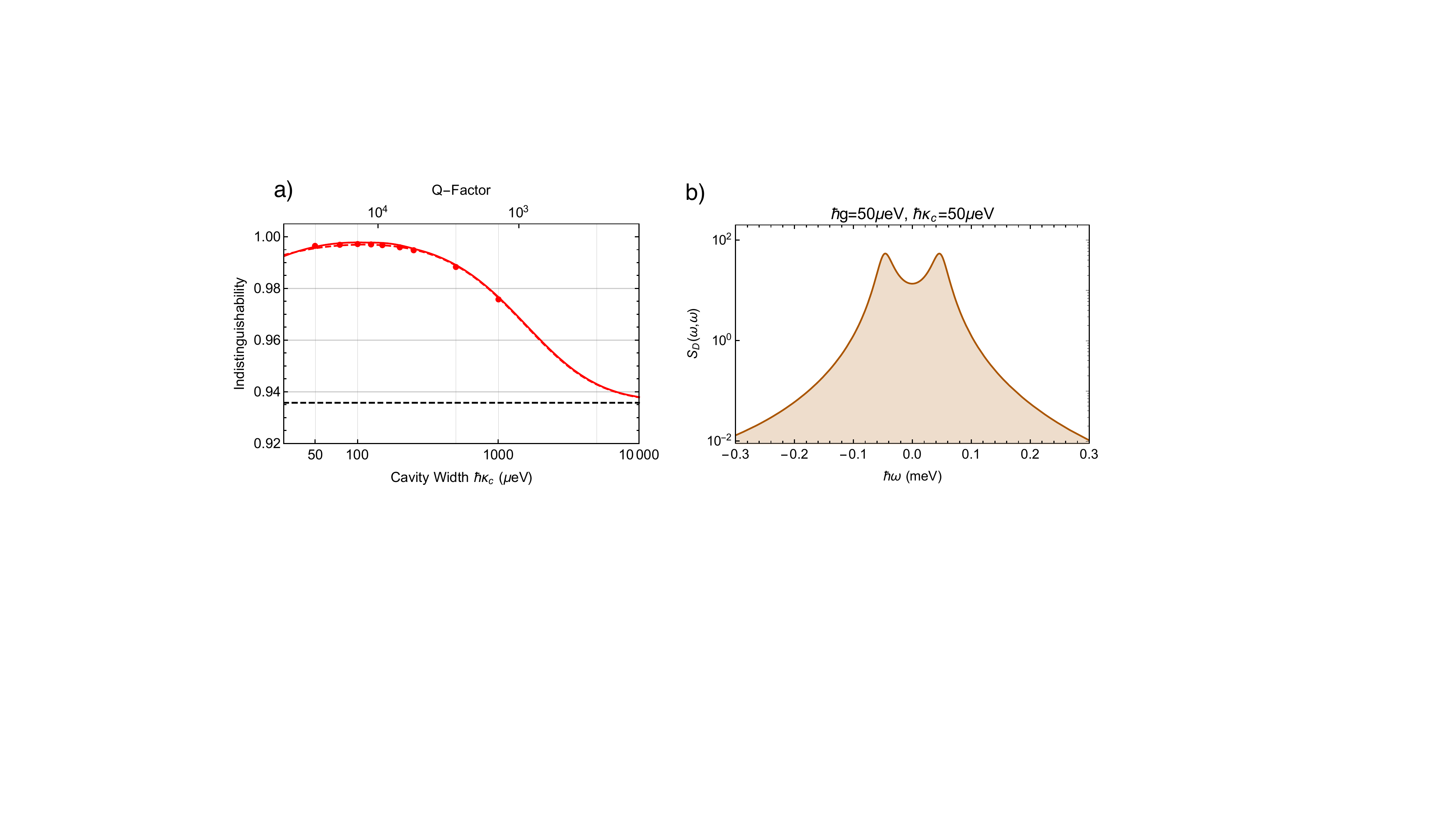}
	\caption{a) Comparison of the indistinguishability calculated from our theory (solid curve), our 
		approximate expression, Eq.~({\ref{IWithSB}}) (dashed curve), and the exact 
		diagonalisation method (plot markers) from Ref.~\cite{PhysRevB.90.035312}, as a function of the cavity width. 
		Part b) shows the detected emission spectra for the parameters indicated. 
		The parameters used are as in Fig.~({\ref{TheoryComparison}}) with $\hbar g = 50~\mu\mathrm{eV}$.}
	\label{comparison}
\end{figure}

With reference to Eqs.~({\ref{WGMEQ}}) and ({\ref{CavMEAd}}), we see that the QD is described by a master equation which takes the same form 
in both the waveguide and cavity cases. The first-order correlation function may then be 
written $|\langle\sigma^\dagger(t+\tau)\sigma(t)\rangle| = \exp\left(-\Gamma_{\mathrm{tot}}t - \frac{1}{2}(\Gamma_{\mathrm{tot}} +2\gamma_{\mathrm{tot}})\tau\right)$. 
From Eq.~(2) in the main text, we write the indistinguishability as
\begin{equation} 
\mathcal{I} = P_D^{-2} \int_{-\infty}^\infty d\omega \int_{-\infty}^\infty d\nu \vert{h(\omega)}\vert^2 \vert{h(\nu)}\vert^2 
\vert S_{\mathrm{ZPL}}(\omega,\nu)+S_{\mathrm{SB}}(\omega,\nu)\vert^2,
\end{equation}
where the function $h(\omega)=1$ in case (i), or describes the filter or cavity in cases (ii) and (iii) respectively.
The contribution from the phonon sideband is purely incoherent, and as such the term involving $S_{\mathrm{SB}}$ may be neglected. 
Making use also of the assumption that the ZPL is unaffected by the filter or cavity we can write 
\begin{equation}
\mathcal{I}\approx P_D^{-2} \int_{-\infty}^\infty d\omega \int_{-\infty}^\infty d\nu \vert S_{\mathrm{ZPL}}(\omega,\nu)\vert^2
=\frac{1}{P_D^2}\frac{4\pi^2 B^4\Gamma_O^2}{\Gamma_{\mathrm{tot}}(\Gamma_{\mathrm{tot}}   + 2\gamma_{\mathrm{tot}})}.
\end{equation}
Using again that the ZPL is unfiltered, we approximate $P_D \approx P_{\mathrm{ZPL}} + \mathcal{F}P_{\mathrm{SB}}$, where 
$P_{\mathrm{ZPL}} = \int_{-\infty}^\infty S_{\mathrm{ZPL}}(\omega,\omega)d\omega= 2\pi B^2\Gamma_O\Gamma^{-1}_{\mathrm{tot}}$ is the power in the ZPL, 
and $P_{\mathrm{SB}} = \int_{-\infty}^\infty S_{\mathrm{SB}}(\omega,\omega)d\omega=2\pi\Gamma_O\Gamma_{\mathrm{tot}}^{-1}(1-B^2)$
is the power in the phonon sideband, while $\mathcal{F} = P_{\mathrm{SB}}^{-1} \int_{-\infty}^\infty \vert h(\omega)\vert^2S_{\mathrm{SB}}(\omega,\omega)d\omega$ 
is the fraction of the phonon sideband not removed by the filter or cavity.
Piecing this together, we obtain:
\begin{equation}
\mathcal{I} = \frac{\Gamma_{\mathrm{tot}}}{\Gamma_{\mathrm{tot}} + 2\gamma_{\mathrm{tot}}}\left(\frac{B^2}{B^2 + \mathcal{F}[1-B^2]}\right)^2,
\label{IWithSB}
\end{equation}
which is Eq.~(4) in the main text. 
The analytic expressions for the efficiencies quoted in the main text are obtained 
with the same approximation $P_D\approx P_{\mathrm{ZPL}}+\mathcal{F}P_{\mathrm{SB}}$ quoted above.

In order to benchmark our formalism, in Fig.~({\ref{comparison}}-a) we compare our full theory where the cavity 
is not adiabatically eliminated (solid curve), the adiabatic approximation expressed in Eq.~({\ref{IWithSB}}) (dashed curve), 
and the exact diagonalisation method of Ref.~\cite{PhysRevB.90.035312} (plot markers). We see that both the full theory and 
the approximation above match the numerically exact results excellently, and both tend towards the expected 
value of $\mathcal{I}=B^4$ in the limit $\kappa_f \to \infty$. Finally, in Fig.~({\ref{comparison}}-b) we 
show the corresponding detected emission spectrum for the parameters indicated, which displays vacuum Rabi splitting. 
This demonstrates that our full theory, as well as our analytic expressions remain valid at the onset of the 
strong coupling regime. 

\providecommand{\noopsort}[1]{}\providecommand{\singleletter}[1]{#1}%

\end{document}